# A New Achievable Rate Region for the Cognitive Radio Channel

Ghosheh Abed Hodtani

Department of Electrical Engineering, Ferdowsi University of Mashhad, Mashhad, Iran

ghodtani@gmail.com

**Abstract**. Considering a general input distribution, using Gel'fand-Pinsker full binning scheme and the Han-Kobayashi (HK) jointly decoding strategy, we obtain a new achievable rate region for the cognitive radio channel (CRC) and then derive a simplified description for the region, by a combination of Cover superposition coding, binning scheme and the HK decoding technique. Our rate region (i) has an interesting interpretation, i.e., any rate in the region, as expected, has generally three additional terms in comparison with the KH region for the interference channel (IC): one term due to the input correlation, the other term due to binning scheme and the third term due to the interference dependent on the inputs, (ii) is really a generalization of the HK region for the IC to the CRC by the use of binning scheme, and as a result of this generalization we see that different versions of our region for the CRC are reduced to different versions of the previous results for the IC, and (iii) is a generalized and improved version of previous results ,i.g., the Devroye-Mitran-Tarokh (DMT) region.

## I. INTRDUCTION

Cognitive radio [1] is one of the most promising technologies to improve spectrum utilization and hence to increase spectrum efficiency in wireless communications, where the traditional fixed spectrum policy is inefficient. In cognitive radio either a network or a wireless node (secondary user) at first senses the environment (primary user) and then changes its transmission or reception parameters in order to adapt to the environment and to efficiently communicate with licensed or unlicensed users. Therefore, in cognitive radio channel (CRC) we have sender-receiver pairs as in the interference channel (IC).

Here we consider only two user CRC and study it from only information theoretical point of view. The IC has been studied information-theoretically in detail (for example, see [2],[3] and references therein). In [4]-[6], the fundamental theoretical questions on the limits of cognitive radio technology are investigated. Specifically, in [7] by directly applying the HK jointly decoding strategy [8] and binning scheme [9], a rate region has been obtained. Also, in the literature we have a rate region [10], outer bounds [11], special capacity region [12],[13] and a simplified description [14] for the rate in [7].

In this paper, we look at the CRC from a new point of view,i.e., the CRC includes the IC and the results for the IC can be generally derived from the studies of the CRC. Aiming at this viewpoint, by considering a general input distribution, using binning scheme and the HK jointly decoding strategy, we obtain a new achievable rate region for the CRC and then derive a simplified description for the region. Our rate region (i) has an interesting interpretation, i.e., any rate in the region, as expected, has generally three additional terms in comparison with the HK region [8] for the IC: one term due to the input correlation, the other term due to binning scheme and the third term due to the interference dependent on the inputs, (ii) is really a generalization of the HK region for the IC to the CRC by the use of binning scheme, and as a result of this generalization we see that different versions of our region for the CRC are reduced to different versions of the previous results for the IC, and (iii) is a generalized and improved version of previous results ,i.g., the Devroye-Mitran-Tarokh (DMT) region [7].

The remainder of the paper is as follows. In section II, we define the IC, the CRC and their modified versions. In section III, we remember briefly some of the previous works which are related directly to our work. In section IV, we explain our main results (four theorems: two theorems for our rate region and two other theorems for its simplified descriptions). In section V, the corollaries of our theorems are explained: first, we prove that the previous works for the IC are special cases of our theorems; second, we show that the previous works for the CRC are included in our region. Finally, we have a conclusion in section VI.

## II. DEFINITIONS

We denote random variables by $X_1, X_2, Y_1, \cdots$ with values $x_1, x_2, y_1, \cdots$ in finite sets $\mathcal{X}_1, \mathcal{X}_2, \mathcal{Y}_1, \cdots$ respectively; n-tuple vectors of $X_1, X_2, Y_1, \cdots$ are denoted with $\boldsymbol{x_1}, \boldsymbol{x_2}, \boldsymbol{y_1}, \cdots$. We use the symbol $A_\varepsilon^n(X_1, X_2, \cdots, X_l)$ to indicate the set of $\varepsilon$-typical n-sequences $(\boldsymbol{x_1}, \boldsymbol{x_2}, \cdots, \boldsymbol{x_l})$ [15].

Interference Channel (IC)

A discrete and memoryless IC $(\mathcal{X}_1 \times \mathcal{X}_2, p(y_1y_2|x_1x_2), \mathcal{Y}_1 \times \mathcal{Y}_2)$ consists of two sender-receiver pairs $(X_1 \to Y_1$ and $X_2 \to Y_2)$ in Fig.1, where $\mathcal{X}_1, \mathcal{X}_2$ are two finite input alphabet sets; $\mathcal{Y}_1, \mathcal{Y}_2$ are two finite output alphabet sets, and $p(y_1y_2|x_1x_2)$ is a conditional channel probability of $(y_1, y_2) \in \mathcal{Y}_1 \times \mathcal{Y}_2$ given $(x_1, x_2) \in \mathcal{X}_1 \times \mathcal{X}_2$. The nth extension of the channel is:

$$p(\mathbf{y_1 y_2}|\mathbf{x_1 x_2}) = \prod_{i=1}^{n} p(y_{1i}y_{2i}|x_{1i}x_{2i})$$

A code $(n, M_1 = \lfloor 2^{nR_1} \rfloor, M_2 = \lfloor 2^{nR_2} \rfloor, \varepsilon)$ is a collection of $M_1$ codewords $\mathbf{x_{1i}} \in \mathcal{X}_1^n, i \in \mathcal{M}_1$; $M_2$ codewords $\mathbf{x_{2j}} \in \mathcal{X}_2^n, j \in \mathcal{M}_2$; two decoding functions $g_1: \mathbf{y_1} \to \mathcal{M}_1$, $g_2: \mathbf{y_2} \to \mathcal{M}_2$; and the average error probabilities at the receivers $(P_{e_1}^n, P_{e_2}^n)$ are defined conveniently [8],[17].

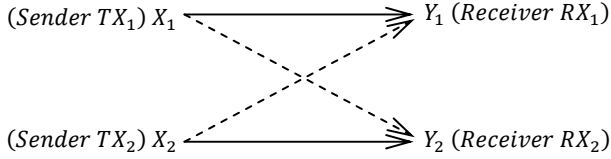 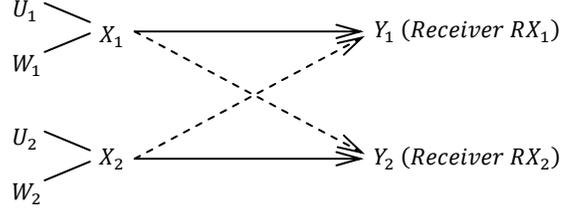

Fig.1 Interference channel       Fig.2 Modified interference channel

A pair $(R_1, R_2)$ of non-negative real values is called an achievable rate if there exists a sequence of codes such that under some decoding scheme, $\max(P_{e_1}^n, P_{e_2}^n) < \varepsilon$.

The capacity region of the IC is the set of all achievable rates.

Modified interference channel

As in [8], a modified IC (Fig.2), models two senders communicating both private and common message to two receivers; where the information conveying role of the channel inputs $X_1, X_2$ is transferred to some fictitious inputs $U_1, W_1, U_2, W_2$, so that the channel behaves like a channel $U_1, W_1, U_2, W_2 \to Y_1Y_2$.

Auxiliary random variables $W_1$ and $W_2$ represent the public message to be sent from $TX_1$ to $(RX_1, RX_2)$ with the rate $T_1$ and from $TX_2$ to $(RX_1, RX_2)$ with the rate $T_2$, respectively. Similarly, $U_1$ and $U_2$ are the private message to be sent from $TX_1$ to $RX_1$ with the rate $S_1$ and from $TX_2$ to $RX_2$ with the rate $S_2$, respectively. Also, as in [8], $Q \in \mathcal{Q}$ is a time sharing random variable whose n-sequences $\mathbf{q} = (q_1, q_2, \cdots, q_n)$ are generated independently of the messages. The n-sequences $\mathbf{q}$ are given to both senders and receivers.

An $(n, \lfloor 2^{nT_1} \rfloor, \lfloor 2^{nS_1} \rfloor, \lfloor 2^{nT_2} \rfloor, \lfloor 2^{nS_2} \rfloor, \varepsilon)$ code for the modified IC (Fig.2) consists of $\lfloor 2^{nT_1} \rfloor$ codewords $\mathbf{w_1}(j)$, $\lfloor 2^{nS_1} \rfloor$ codewords $\mathbf{u_1}(l)$ for $TX_1$; and $\lfloor 2^{nT_2} \rfloor$ codewords $\mathbf{w_2}(m)$, $\lfloor 2^{nS_2} \rfloor$ codewords $\mathbf{u_2}(k)$ for $TX_2$ ; $j \in \{1, \cdots, 2^{nT_1}\}$, $l \in \{1, \cdots, 2^{nS_1}\}$, $m \in \{1, \cdots, 2^{nT_2}\}$, $k \in \{1, \cdots, 2^{nS_2}\}$, such that the maximum of the conveniently defined average probabilities of decoding error $(P_{e_1}^n, P_{e_2}^n)$ is less than $\varepsilon$.

A quadruple $(T_1, S_1, T_2, S_2)$ of non-negative real numbers is achievable for the modified IC (and hence, $(R_1 = S_1 + T_1, R_2 = S_2 + T_2)$ is achievable rate for the IC) if there exists a sequence of codes such that the maximum of average error probabilities under some decoding scheme is less than $\varepsilon$. An achievable region for the modified IC is the closure of a subset of the positive region $R^4$ of achievable rate quadruples $(T_1, S_1, T_2, S_2)$.

Cognitive radio channel

A cognitive radio channel (Fig.3) is defined to be an IC (Fig.1) in which $TX_2$ is non-causally aware of the message to be sent by $TX_1$.

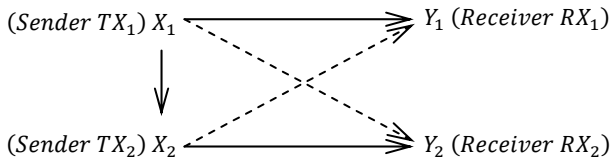 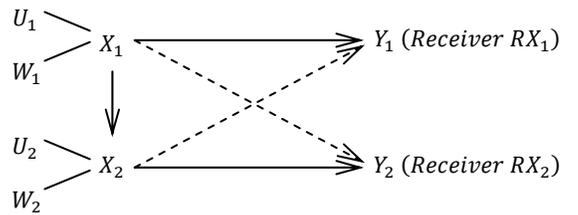

Fig.3 Cognitive radio channel       Fig. 4 Modified cognitive radio channel

Modified cognitive radio channel

A modified CRC (Fig.4) is a CRC, in which we consider the inputs $U_1W_1U_2W_2$ instead of $X_1X_2$ , and $X_1$ (and hence $U_1W_1$) is known non-causally at $TX_2$.

A code and achievable rate for modified CRC (Fig.4) are defined the same as for modified IC (Fig.2), but with taking account of the dependence of $U_2W_2$ on $U_1W_1$.

General distributions on auxiliary, input and output channel random variables for the IC and the CRC

Therefore, we can consider auxiliary random variables $Q, U_1, W_1, U_2, W_2$, defined on arbitrary finite sets $\mathcal{Q}, \mathcal{U}_1, \mathcal{W}_1, \mathcal{U}_2, \mathcal{W}_2$, respectively; $X_1$ and $X_2$ defined on the input alphabet sets $\mathcal{X}_1, \mathcal{X}_2$, and $Y_1, Y_2$, defined on the output alphabet sets $\mathcal{Y}_1$ and $\mathcal{Y}_2$. Let $Z = (QU_1W_1U_2W_2X_1X_2Y_1Y_2)$ and let $\mathcal{P}_{IC}$ be the set of all distributions of the form (for Fig.2): (hereafter, for brevity, let $p(qu_1w_1u_2w_2x_1x_2y_1y_2) = p(\ )$)

$$p(\ ) = p(q)p(u_1w_1|q)p(u_2w_2|q)p(x_1x_2|qu_1u_2w_1w_2)p(y_1y_2|x_1x_2) \qquad (1),$$

and $\mathcal{P}_{CRC}$ be the set of all distributions of the form (for Fig.4):

$$p(\ ) = p(q)p(u_1w_1|q)p(u_2w_2|qw_1u_1)p(x_1x_2|qu_1w_1u_2w_2)p(y_1y_2|x_1x_2) \qquad (2).$$

### III. RELATED WORKS

**For the IC**

Han and Kobayashi [8] considered the general distribution (1) in the following form:

$$p(\ ) = p(q)p(u_1|q)p(w_1|q)p(u_2|q)p(w_2|q)p(x_1x_2|qu_1u_2w_1w_2)p(y_1y_2|x_1x_2) \qquad (3),$$

and derived the best achievable rate region known to date. This region is described in terms of quadruple $(T_1, S_1, T_2, S_2)$ as in theorem 3.1, [8] or theorem 1, [2] and in terms of $(R_1, R_2)$ as in theorem B, [16] or its amended version in theorem 2, [2]. The simplified description of the HK region known as the Chong-Motani-Garg (CMG) region is investigated [17] or [2] (theorems 4 and 5) with superposition coding and the HK decoding strategy, using the input distribution below (in Fig.2) :

$$p(qw_1x_1w_2x_2) = p(q)p(w_1|q)p(x_1|qw_1)p(w_2|q)p(x_2|qw_2) \qquad (4)$$

**For the CRC**

Devroye-Mitran-Tarokh (DMT) obtained an achievable rate region for the CRC (Fig.4) with a special case of the general distribution (2) in the following form:

$$p(\ ) = p(q)p(w_1|q)p(u_1|q)p(w_2|qu_1w_1)p(u_2|qu_1w_1)p(x_1x_2|qu_1u_2w_1w_2)p(y_1y_2|x_1x_2) \qquad (5).$$

In accordance with our definitions and notations in section II, The DMT region is described as follows.

**Theorem 1** (theorem 1, [7], the inequalities 6-21)**:** For the modified CRC (Fig.4), let $Z = (QU_1W_1U_2W_2X_1X_2Y_1Y_2)$ and let $\mathcal{P}_{CRC}^{DMT}$ be the set of all distributions of the form (5). For any $Z \in \mathcal{P}_{CRC}^{DMT}$ let $S_{CRC}^{DMT}(Z)$ be the set of all quadruples $(T_1, S_1, T_2, S_2)$ such that

$$S_1 \leq I(W_1W_2; U_1|Q) + I(Y_1; U_1|QW_1W_2) = a_1, \qquad (6\text{-}1),$$

$$T_1 \leq I(U_1W_2; W_1|Q) + I(Y_1; W_1|QW_2U_1) = b_1, \qquad (6\text{-}2),$$

$$T_2 \leq I(Y_1; W_2|QU_1W_1) = c_1, \qquad (6\text{-}3),$$

$$S_1 + T_1 \leq I(W_2; W_1U_1|Q) + I(Y_1; U_1W_1|QW_2) = d_1, \qquad (6\text{-}4),$$

$$S_1 + T_2 \leq I(Y_1; U_1W_2|QW_1) + I(U_1W_2; W_1|Q) - I(W_2; W_1U_1|Q) = e_1, \qquad (6\text{-}5),$$

$$T_1 + T_2 \leq I(W_1W_2; U_1|Q) + I(Y_1; W_1W_2|QU_1) - I(W_2; W_1U_1|Q) = f_1, \qquad (6\text{-}6),$$

$$S_1 + T_1 + T_2 \leq I(Y_1; U_1W_1W_2|Q) - I(W_2; W_1U_1|Q) = g_1, \qquad (6\text{-}7),$$

$$S_2 \leq I(U_2; W_1W_2|Q) + I(Y_2; U_2|QW_1W_2) - I(U_2; U_1W_1|Q) = a_2, \qquad (6\text{-}8),$$

$$T_2 \leq I(W_2; W_1U_2|Q) + I(Y_2; W_2|QW_1U_2) - I(W_2; W_1U_1|Q) = b_2, \qquad (6\text{-}9),$$

$$T_1 \leq I(W_2U_2; W_1|Q) + I(Y_2; W_1|QW_2U_2) = c_2, \qquad (6\text{-}10),$$

$$S_2 + T_2 \leq I(W_2U_2; W_1|Q) + I(Y_2; U_2W_2|QW_1) - I(W_2; U_1W_1|Q) = d_2, \qquad (6\text{-}11),$$

$$S_2 + T_1 \leq I(W_1U_2; W_2|Q) + I(Y_2; U_2W_1|QW_2) - I(U_2; U_1W_1|Q) = e_2, \qquad (6\text{-}12),$$

$$T_1 + T_2 \leq I(U_2; W_1W_2|Q) + I(Y_2; W_1W_2|QU_2) - I(W_2; U_1W_1|Q) = f_2, \qquad (6\text{-}13),$$

$$S_2 + T_1 + T_2 \leq I(Y_2; U_2W_1W_2|Q) - I(W_2; W_1U_1|Q) - I(U_2; U_1W_1|Q) = g_2 \qquad (6\text{-}14),$$

$$S_i \geq 0, \quad T_i \geq 0, \quad i = 1,2$$

then any element of the closure of $\bigcup_{Z \in \mathcal{P}_{CRC}^{DMT}} S_{CRC}^{DMT}(Z)$ is achievable.

**Proof**. Refer to [7].

Also, Rini-Tuninetti-Devroye (RTD) [18], for the CRC in Fig.4, considered $U_1 = (U_{1a}, U_{1b})$, $Q = \emptyset$ and the following distribution (as seen in outline of proof of theorem 1 in [18]):

$$p(\ ) = p(w_1)\,p(u_{1a}|w_1)\,p(w_2|w_1u_{1a})\,p(u_2|w_1w_2u_{1a})\,p(u_{1b}|w_1w_2u_{1a}u_2)\,p(x_1|u_{1a}w_1)\,p(x_2|w_1w_2u_{1a}u_{1b}u_2) \qquad (7),$$

and using superposition coding and partial binning scheme, obtained a rate region which can be written according to our notations in section II as follows.

**Theorem 2** (theorem 1, [18])**:** For the modified CRC (Fig.4), let $Z = (QU_{1a}U_{1b}W_1U_2W_2X_1X_2Y_1Y_2)$ and let $\mathcal{P}_{CRC}^{RTD}$ be the set of all distributions of the form (7). For any $Z \in \mathcal{P}_{CRC}^{RTD}$ let $S_{CRC}^{RTD}(Z)$ be the set of all quintuples $(T_1, S_{1a}, S_{1b}, T_2, S_2)$ such that

$$T_1 + T_2 + S_{1a} + S_{1b} \leq I(Y_1; W_2W_1U_{1a}U_{1b}) - I(U_2; U_{1b}|W_1W_2U_{1a}) \quad (8\text{-}1),$$

$$T_2 + S_{1a} + S_{1b} \leq I(Y_1; W_2U_{1a}U_{1b}|W_1) - I(U_2; U_{1b}|W_1W_2U_{1a}) \quad (8\text{-}2),$$

$$S_{1a} + S_{1b} \leq I(Y_1; U_{1a}U_{1b}|W_1W_2) + I(W_2; U_{1a}|W_1) - I(U_2; U_{1b}|W_1W_2U_{1a}) \quad (8\text{-}3),$$

$$T_2 + S_{1b} \leq I(Y_1; W_2U_{1b}|W_1U_{1a}) - I(U_2; U_{1b}|W_1W_2U_{1a}) \quad (8\text{-}4),$$

$$S_{1b} \leq I(Y_1; U_{1b}|W_1W_2U_{1a}) + I(W_2; U_{1a}|W_1) - I(U_2; U_{1b}|W_1W_2U_{1a}) \quad (8\text{-}5),$$

$$T_1 + T_2 + S_2 \leq I(Y_2; W_1W_2U_2) - I(W_2; U_{1a}|W_1) - I(U_2; U_{1a}|W_1W_2) \quad (8\text{-}6),$$

$$T_2 + S_2 \leq I(Y_2; W_2U_2|W_1) - I(W_2; U_{1a}|W_1) - I(U_2; U_{1a}|W_1W_2) \quad (8\text{-}7),$$

$$S_2 \leq I(Y_2; U_2|W_1W_2) - I(U_2; U_{1a}|W_1W_2) \quad (8\text{-}8),$$

then any element of the closure of $\bigcup_{Z \in \mathcal{P}_{CRC}^{RTD}} S_{CRC}^{RTD}(Z)$ is achievable.

**Proof.** Refer to [18].

## IV. MAIN RESULTS

Our motivation for this study is the facts that (i) the best achievable rate region for broadcast channels has been obtained by using the binning scheme and considering a general input distribution [19] and,(ii) multiple access channels have been studied first for independent inputs [20],[21], then for specially correlated inputs [22] and ultimately for arbitrarily correlated inputs [23]. Having originated from this motivation, we study the CRC in Fig.4, with general input distribution (2) in the form of:

$$p(\ ) = p(q)p(w_1|q)p(u_1|qw_1)p(w_2|qu_1w_1)p(u_2|qu_1w_1w_2) \quad (9),$$

and by using the binning scheme and the HK decoding strategy, we obtain a rate region which is described in terms of quadruples $(T_1, S_1, T_2, S_2)$ with fourteen relations (theorem 3) and can be transformed by virtue of the Fourier-Motzkin algorithm into the rate pair $(R_1, R_2)$ including twenty inequalities (theorem 4). And then, we derive a simplified description for our region, by appropriately making use of superposition coding, binning scheme and jointly decoding, in terms of quadruples $(T_1, S_1, T_2, S_2)$ and the rate pair $(R_1, R_2)$ (theorems 5 and 6).

**New rate region for the CRC**

**Theorem 3 :** For the modified CRC (Fig.4), let $Z = (QU_1W_1U_2W_2X_1X_2Y_1Y_2)$ and let $\mathcal{P}_{CRC}^{Hod}$ be the set of all distributions of the form (9). For any $Z \in \mathcal{P}_{CRC}^{Hod}$ let $S_{CRC}^{Hod}(Z)$ be the set of all quadruples $(T_1, S_1, T_2, S_2)$ such that

$$S_1 \leq I(W_2; U_1W_1|Q) + I(Y_1; U_1|QW_1W_2) = A_1 \quad (10\text{-}1),$$

$$T_1 \leq I(U_1; W_1|Q) + I(W_2; U_1W_1|Q) + I(Y_1; W_1|QW_2U_1) = B_1 \quad (10\text{-}2),$$

$$T_2 \leq I(U_1; W_1|Q) + I(Y_1; W_2|QU_1W_1) = C_1 \quad (10\text{-}3),$$

$$S_1 + T_1 \leq I(W_2; W_1U_1|Q) + I(Y_1; U_1W_1|QW_2) = D_1 \quad (10\text{-}4),$$

$$S_1 + T_2 \leq I(Y_1; U_1W_2|QW_1) = E_1 \quad (10\text{-}5),$$

$$T_1 + T_2 \leq I(U_1; W_1|Q) + I(Y_1; W_1W_2|QU_1) = F_1 \quad (10\text{-}6),$$

$$S_1 + T_1 + T_2 \leq I(Y_1; U_1W_1W_2|Q) = G_1 \quad (10\text{-}7),$$

$$S_2 \leq I(U_2; W_2|Q) + I(W_2U_2; W_1|Q) + I(Y_2; U_2|QW_1W_2) - I(U_2; U_1W_1W_2|Q) = A_2 \quad (10\text{-}8),$$

$$T_2 \leq I(U_2; W_2|Q) + I(W_1; W_2U_2|Q) + I(Y_2; W_2|QW_1U_2) - I(W_2; W_1U_1|Q) = B_2 \quad (10\text{-}9),$$

$$T_1 \leq I(U_2; W_2|Q) + I(W_2U_2; W_1|Q) + I(Y_2; W_1|QW_2U_2) = C_2 \quad (10\text{-}10),$$

$$S_2 + T_2 \leq I(U_2; W_2|Q) + I(W_2U_2; W_1|Q) + I(Y_2; U_2W_2|QW_1) - I(W_2; U_1W_1|Q) - I(U_2; U_1W_1W_2|Q) = D_2 \quad (10\text{-}11),$$

$$S_2 + T_1 \leq I(U_2; W_2|Q) + I(W_2U_2; W_1|Q) + I(Y_2; U_2W_1|QW_2) - I(U_2; U_1W_1W_2|Q) = E_2 \quad (10\text{-}12),$$

$$T_1 + T_2 \leq I(U_2; W_2|Q) + I(W_2U_2; W_1|Q) + I(Y_2; W_1W_2|QU_2) - I(W_2; U_1W_1|Q) = F_2 \quad (10\text{-}13),$$

$$S_2 + T_1 + T_2 \leq I(U_2; W_2|Q) + I(W_1; U_2W_2|Q) + I(Y_2; U_2W_1W_2|Q) - I(W_2; W_1U_1|Q) - I(U_2; U_1W_1W_2|Q) = G_2 \quad (10\text{-}14),$$

$S_i \geq 0$ , $T_i \geq 0$ , $i = 1,2$

then any element of the closure of $\bigcup_{Z\in\mathcal{P}_{CRC}^{Hod}} S_{CRC}^{Hod}(Z)$ is achievable.

**Proof**. Refer to Appendix A.

We can interpret the above region as follows.

**Interpretation of theorem 3:**

We have allowed the input auxiliary variables to be dependent; used the binning scheme conveniently and derived an improved rate region. Specifically, we have added generally three terms as seen in (10-1)-(10-14) to every rate in the HK region for the IC yielding to the rate region for the CRC:

One positive term is due to the input correlation $I(U_1; W_1|Q)$ or $I(U_2; W_2|Q)$; the second term is related to the interference dependent on the inputs at the corresponding receiver $I(W_2; U_1W_1|Q)$ or $I(W_1; U_2W_2|Q)$, originating from the HK jointly decoding strategy. The third additional term (negative term $-I(W_2; U_1W_1|Q)$ or $-I(U_2; U_1W_1W_2|Q)$) is a result of the binning scheme.

**Remark**. If we consider a common message $W_0$ with the rate $T_0$ for the transmitters in the CRC (Fig.4), we will have the rates $(S_1, T_1, T_0, T_2)$, $(S_2, T_2, T_0, T_1)$ for the first and second transmitters, respectively. For this more general case, we can prove that the quintuple $(S_1, T_2, T_1, T_0, T_2)$ satisfies 30 inequalities, instead of 14 inequalities in theorem 3. Theorem 3 is a special case of this general situation.

Now, we transform the above region in theorem 3 into the rate pair $(R_1 = S_1 + T_1, R_2 = S_2 + T_2)$ region by the Fourier-Motzkin elimination technique and find a general theorem for the IC and CRC the corollaries of which will be explained in section V.

**Theorem 4.** The $S_{CRC}^{Hod}$ region in theorem 3 can be described as $\mathcal{R}_{Hod}$ being the set of $(R_1, R_2)$ satisfying:

$$R_1 \leq D_1 \quad (11\text{-}1),$$
$$R_1 \leq A_1 + C_2 \quad (11\text{-}2),$$
$$R_1 \leq G_1 \quad (11\text{-}3),$$
$$R_1 \leq F_2 + A_1 \quad (11\text{-}4),$$
$$R_1 \leq C_2 + E_1 \quad (11\text{-}5),$$
$$R_1 \leq B_1 + E_1 \quad (11\text{-}6),$$
$$R_1 \leq E_2 + A_1 \quad (11\text{-}7),$$
$$R_1 \leq F_2 + E_1 \quad (11\text{-}8),$$
$$R_2 \leq D_2 \quad (11\text{-}9),$$
$$R_2 \leq A_2 + C_1 \quad (11\text{-}10),$$
$$R_2 \leq E_1 + A_2 \quad (11\text{-}11),$$
$$R_1 + R_2 \leq E_1 + E_2 \quad (11\text{-}12),$$
$$R_1 + R_2 \leq G_2 + A_1 \quad (11\text{-}13),$$
$$R_1 + R_2 \leq G_1 + A_2 \quad (11\text{-}14),$$
$$R_1 + R_2 \leq B_1 + E_1 + A_2 \quad (11\text{-}15),$$
$$R_1 + R_2 \leq E_1 + G_2 \quad (11\text{-}16),$$
$$2R_1 + R_2 \leq G_1 + E_2 + A_1 \quad (11\text{-}17),$$
$$2R_1 + R_2 \leq F_2 + E_2 + 2A_1 \quad (11\text{-}18),$$
$$2R_2 + R_1 \leq F_1 + E_1 + 2A_2 \quad (11\text{-}19),$$
$$2R_2 + R_1 \leq E_1 + G_2 + A_2 \quad (11\text{-}20),$$

where the bound constants $A_i$, $B_i$, $C_i$, $D_i$, $E_i$, $F_i$, $G_i$, $i = 1,2$ are the same as in theorem 3.

**Proof**. Refer to Appendix B.

**Simplified description of our rate region for the CRC**

In this subsection, we reduce the number of auxiliary random variables in (9), by using superposition coding, i.e., we transform $QW_1U_1$ to $QW_1X_1$ and $QW_2U_2$ to $QW_2X_2$ such that the message conveyed by $U_1(U_2)$ are superimposed over $QW_1(QW_2)$ with $X_1(X_2)$, respectively; and totally we exploit the following distribution resulted of (9):

$$p(qw_1x_1w_2x_2) = p(q)p(w_1|q)p(x_1|qw_1)p(w_2|qw_1x_1)p(x_2|qw_2w_1x_1) \quad (12),$$

and combining superposition coding, binning scheme and jointly decoding, we reach to the simplified description below.

**Theorem 5**. For the modified CRC (Fig.4), let $Z_1 = (QW_1W_2X_1X_2Y_1Y_2)$ and let $\mathcal{P}_{CRC}^{Hod-1}$ be the set of all distributions of the form (12). For any $Z_1 \in \mathcal{P}_{CRC}^{Hod-1}$ let $S_{CRC}^{Hod-1}(Z_1)$ be the set of all quadruples $(T_1, S_1, T_2, S_2)$ such that

$$S_1 \leq A_1 \qquad (13\text{-}1),$$
$$S_1 + T_1 \leq D_1 \qquad (13\text{-}2),$$
$$S_1 + T_2 \leq E_1 \qquad (13\text{-}3),$$
$$S_1 + T_1 + T_2 \leq G_1 \qquad (13\text{-}4),$$
$$S_1 \geq 0, \qquad T_1 \geq 0,$$

and the inequalities (13-5)-(13-8) similarly to (13-1)-(13-4) with the indices 1 and 2 swapped. Then any element of the closure of $\bigcup_{Z_1 \in \mathcal{P}_{CRC}^{Hod-1}} S_{CRC}^{Hod-1}(Z_1)$ is achievable, where in (13-1)-(13-8) we have:

$$A_1 = I(Y_1; U_1|W_1W_2Q) + I(W_2; W_1U_1|Q) = I(Y_1; X_1|W_1W_2Q) + I(W_2; X_1|Q) \qquad (14\text{-}1),$$
$$D_1 = I(Y_1; U_1W_1|W_2Q) + I(W_2; W_1U_1|Q) = I(Y_1; X_1|W_2Q) + I(W_2; X_1|Q) \qquad (14\text{-}2),$$
$$E_1 = I(Y_1; U_1W_2|W_1Q) = I(Y_1; X_1W_2|W_1Q) \qquad (14\text{-}3),$$
$$G_1 = I(Y_1; U_1W_1W_2|Q) = I(Y_1; X_1W_2|Q) \qquad (14\text{-}4),$$
$$A_2 = I(U_2; W_2|Q) + I(W_2U_2; W_1|Q) + I(Y_2; U_2|QW_1W_2) - I(U_2; U_1W_1W_2|Q) =$$
$$I(X_2; W_1|Q) + I(Y_2; X_2|W_1W_2Q) - I(X_2; X_1|QW_2) \qquad (14\text{-}5),$$
$$D_2 = I(U_2; W_2|Q) + I(W_2U_2; W_1|Q) + I(Y_2; U_2W_2|QW_1) - I(W_2; U_1W_1|Q) - I(U_2; U_1W_1W_2|Q) =$$
$$I(Y_2; X_2|W_1Q) - I(W_2; X_1|Q) - I(X_2; X_1|QW_2) + I(X_2; W_1|Q) \qquad (14\text{-}6),$$
$$E_2 = I(U_2; W_2|Q) + I(W_2U_2; W_1|Q) + I(Y_2; U_2W_1|QW_2) - I(U_2; U_1W_1W_2|Q) =$$
$$I(X_2; W_1|Q) + I(Y_2; X_2W_1|QW_2) - I(X_2; X_1|QW_2) \qquad (14\text{-}7),$$
$$G_2 = I(U_2; W_2|Q) + I(W_1; U_2W_2|Q) + I(Y_2; U_2W_1W_2|Q) - I(W_2; W_1U_1|Q) - I(U_2; U_1W_1W_2|Q) =$$
$$I(Y_2; X_2W_1|Q) - I(W_2; X_1|Q) - I(X_2; X_1|QW_2) + I(X_2; W_1|Q) \qquad (14\text{-}8),$$

**Proof**. Refer to Appendix C.

Now, we describe the above region in terms of the rate pair $(R_1, R_2)$.

**Theorem 6.** By using the Fourier-Motzkin algorithm, the above region in theorem 5 (the inequalities 13-1,2…,8) can be transformed into $\mathcal{R}_{Hod-1}$ being the set of $(R_1, R_2)$ satisfying:

$$R_1 \leq D_1 \qquad (15\text{-}1),$$
$$R_1 \leq A_1 + E_2 \qquad (15\text{-}2),$$
$$R_1 \leq G_1 \qquad (15\text{-}3),$$
$$R_2 \leq D_2 \qquad (15\text{-}4),$$
$$R_2 \leq A_2 + E_1 \qquad (15\text{-}5),$$
$$R_1 + R_2 \leq A_1 + G_2 \qquad (15\text{-}6),$$
$$R_1 + R_2 \leq A_2 + G_1 \qquad (15\text{-}7),$$
$$R_1 + R_2 \leq E_1 + E_2 \qquad (15\text{-}8),$$
$$R_1 + R_2 \leq E_1 + G_2 \qquad (15\text{-}9),$$
$$2R_1 + R_2 \leq A_1 + G_1 + E_2 \qquad (15\text{-}10),$$
$$2R_2 + R_1 \leq A_2 + G_2 + E_1 \qquad (15\text{-}11),$$

where $A_i$, $D_i$, $E_i$, $G_i$, i=1,2 are the same as in theorem 5.

**Proof**. Refer to Appendix D.

## V. COROLLARIES OF OUR THEOREMS

In this section, we explain the corollaries of our theorems stated in section IV:

First, we show that different versions of the HK region and its simplified form (the CMG region) are obtained from theorems 3,4 and 5,6, respectively (corollaries 1-4).

Second, we prove that our work for the CRC is an improved and developed version of previous works, i.e. ,the DMT and the RTD regions (corollaries 5-6).

**Corollary 1.**

The HK region for the IC in terms of the quadruple $(T_1, S_1, T_2, S_2)$ or theorem 3.1,[8] is easily obtained from theorem 3 when we assume the independence of $U_i$ from $W_i$, $i = 1,2$ given $Q$ and also, the independence of $U_2$ and $W_2$ from $W_1 U_1$ given $Q$, as in theorem 3.1,[8].

**Corollary 2.**

The HK region for the IC in terms of the rate pair $(R_1, R_2)$ (amended version of theorem B, [16] or theorem 2, [2]) is readily derived from theorem 4: Considering the relations between the terms $A_i$, $B_i$, $C_i$, $D_i$, $E_i$, $F_i$, $G_i$, $i = 1,2$ in the HK region and the independence of $U_i$ from $W_i$, $i = 1,2$ given $Q$ and also, the independence of $U_2$ and $W_2$ from $W_1 U_1$ given $Q$, the inequalities 11-3,…8,11,15,16,18,19 turn out to be redundant and the remaining 9 inequalities describe the HK region.

**Corollary 3.**

Assuming the independence of $w_2$ from $(w_1)x_1$ given $q$ and $x_2$ from $(w_1)x_1$ given $qw_2$ as in (4), theorem 5 is reduced to the CMG region for the IC (lemma 3,[17]) in terms of the quadruple $(T_1, S_1, T_2, S_2)$.

**Corollary 4.**

The CMG region for the IC in terms of $(R_1, R_2)$ (lemma 4,[17]) is readily obtained from theorem 6 as a result of the redundancy of 15-3,9 due to $D_1 \leq G_1$ and $E_2 \leq G_2$ for the bound constants in the CMG region.

**Corollary 5.**

It can be shown equation by equation that our region for the CRC in theorem 3 is an improved version of the DMT region in (6-1)-(6-14). The terms $a_i$, $b_i$, $c_i$, $d_i$, $e_i$, $f_i$, $g_i$, $i = 1,2$ denote the DMT rates in (6-1)-(6-14) and the terms $A_i$, $B_i$, $C_i$, $D_i$, $E_i$, $F_i$, $G_i$, $i = 1,2$ denote the rates in (10-1)-(10-14) of our region. The comparison is as follows.

$$a_1 = A_1 - I(W_2; W_1|Q) \quad , \qquad a_2 = A_2 - I(W_2; W_1|Q) \quad ,$$
$$b_1 = B_1 - I(W_2; U_1|Q) \quad , \qquad b_2 = B_2 - I(W_1; U_2|Q) \quad ,$$
$$c_1 = C_1 - I(U_1; W_1|Q) \quad , \qquad c_2 = C_2 - I(U_2; W_2|Q) \quad ,$$
$$d_1 = D_1 \quad , \qquad d_2 = D_2 - I(U_2; W_2|Q) - I(W_1; U_2|Q) \quad ,$$
$$e_1 = E_1 - I(W_2; U_1|Q) \quad , \qquad e_2 = E_2 - I(W_1; U_2|Q) \quad ,$$
$$f_1 = F_1 - I(W_2; U_1|Q) \quad , \qquad f_2 = F_2 - I(W_1; W_2|Q) \quad ,$$
$$g_1 = G_1 - I(W_2; U_1 W_1|Q) \quad , \qquad g_2 = G_2 - I(U_2; W_2|Q) - I(W_1; W_2 U_2|Q) \quad .$$

**Corollary 6.**

As seen in corollary 5, our region is easily comparable to the DMT region, whereas it is hard to compare the RTD region with the DMT one as seem in [18]. When $U_{1b} = \emptyset$, our region is better than the RTD region and generally the terms of the RTD region in inequalities (8-1)-(8-8) can be partially compared to the corresponding terms in our region as follows.

$$(S_1 + T_1 + T_2)_{RTD} = (S_1 + T_1 + T_2)_{Hod} - I(U_2; U_{1b}|W_1 W_2 U_{1a})$$
$$(S_1)_{RTD} = (S_1)_{Hod} - I(W_2; W_1) - I(W_2; U_{1b}|W_1 U_{1a})$$
$$(S_1 + T_2)_{RTD} = (S_1 + T_2)_{Hod} - I(U_2; U_{1b}|W_1 W_2 U_{1a})$$
$$(S_{1b} + T_2)_{RTD} = (S_{1b} + T_2)_{Hod} - I(U_2; U_{1b}|W_1 W_2 U_{1a})$$
$$(S_{1b})_{RTD} = (S_{1b})_{Hod} - I(W_2; W_1) - I(U_2 W_2; U_{1b}|W_1 U_{1a})$$
$$(S_2 + T_2 + T_1)_{Hod} = (S_2 + T_1 + T_2)_{RTD} - I(W_2 U_2; U_{1b}|W_1 U_{1a})$$
$$(S_2 + T_2)_{Hod} = (S_2 + T_2)_{RTD} - I(W_2 U_2; U_{1b}|W_1 U_{1a})$$
$$(S_2)_{Hod} = (S_2)_{RTD} + I(W_2; W_1) - I(U_2; U_{1b}|W_1 W_2 U_{1a})$$

## V. CONCLUSION

We have studied the cognitive radio channel with a general input distribution and obtained a new rate region, using the binning scheme and the HK jointly decoding strategy. Our region has been shown to have an interesting interpretation, to include different versions of the HK region for the interference channel and to have a developed and improved version in comparison with previous regions.

# APPENDIX A

**Proof of theorem 3**

We prove the achievability of any element of $S_{CRC}^{Hod}(Z)$ for each $Z \in \mathcal{P}_{CRC}^{Hod}$. Fix $Z = (QU_1W_1U_2W_2X_1X_2Y_1Y_2)$.

**Codebook generation:** Consider $n > 0$, some distribution of the form (9) and

$p(u_1|q) = \sum_{w_1} p(w_1|q)p(u_1|qw_1)$

$p(w_2|q) = \sum_{w_1u_1} p(w_1|q)p(u_1|qw_1)p(w_2|qw_1u_1)$

$p(u_2|q) = \sum_{w_1u_1w_2} p(w_1|q)p(u_1|qw_1)p(w_2|qu_1w_1)p(u_2|qu_1w_1w_2)$.

Therefore, by using random binning we can generate the sequences of $\boldsymbol{u_1}, \boldsymbol{w_2}$ and $\boldsymbol{u_2}$ independently of $\boldsymbol{w_1}, (\boldsymbol{u_1w_1})$ and $(\boldsymbol{u_1w_1w_2})$, respectively.

So,

1. generate a n-sequence $\boldsymbol{q}$, i.i.d. according to $\prod_{i=1}^{n} p(q_i)$, and for the codeword $\boldsymbol{q}$:
2. Generate $\lfloor 2^{nT_1} \rfloor$ conditionally independent codewords $\boldsymbol{w_1}(j)$, $j \in \{1,2,\cdots,\lfloor 2^{nT_1} \rfloor\}$ according to $\prod_{i=1}^{n} p(w_{1i}|q_i)$.
3. Generate $\lfloor 2^{nS_1} \rfloor$ n-sequence $\boldsymbol{u_1}(l)$, $l \in \{1,\cdots,\lfloor 2^{nS_1} \rfloor\}$, i.i.d. according to $\prod_{i=1}^{n} p(u_{1i}|q_i)$ and throw them randomly into $\lfloor 2^{nS_1} \rfloor$ bins such that the sequence $\boldsymbol{u_1}(l)$ in bin $b_1$ is denoted as $\boldsymbol{u_1}(b_1, l)$, $b_1 \in \{1,\cdots,\lfloor 2^{nS_1} \rfloor\}$.
4. Generate $\lfloor 2^{nt_2} \rfloor$ n-sequence $\boldsymbol{w_2}(m)$, $m \in \{1,\cdots,\lfloor 2^{nt_2} \rfloor\}$, i.i.d. according to $\prod_{i=1}^{n} p(w_{2i}|q_i)$ and throw them randomly into $\lfloor 2^{nT_2} \rfloor$ bins such that the sequence $\boldsymbol{w_2}(m)$ in bin $b_2$ is denoted as $\boldsymbol{w_2}(b_2, m)$, $b_2 \in \{1,\cdots,\lfloor 2^{nT_2} \rfloor\}$.
5. Generate $\lfloor 2^{nS_2} \rfloor$ n-sequence $\boldsymbol{u_2}(k)$, $k \in \{1,\cdots,\lfloor 2^{nS_2} \rfloor\}$, i.i.d. according to $\prod_{i=1}^{n} p(u_{2i}|q_i)$ and throw them randomly into $\lfloor 2^{nS_2} \rfloor$ bins such that the sequence $\boldsymbol{u_2}(k)$ in bin $b_3$ is denoted as $\boldsymbol{u_2}(b_3, k)$, $b_3 \in \{1,\cdots,\lfloor 2^{nS_2} \rfloor\}$.

**Encoding:** The messages of $(\boldsymbol{u_1}, \boldsymbol{w_1})$ of $(\mathcal{U}_1^n, \mathcal{W}_1^n)$ and $(\boldsymbol{u_2}, \boldsymbol{w_2})$ of $(\mathcal{U}_2^n, \mathcal{W}_2^n)$ sets are mapped into $x_1$ and $x_2$, respectively through deterministic or random encoding functions. The sender $TX_1$ to send $(j, b_1)$, knowing $\boldsymbol{q}$ looks for $\boldsymbol{w_1}(j)$ and finds a sequence $\boldsymbol{u_1}(l)$ in bin $b_1$ such that $(\boldsymbol{q}, \boldsymbol{w_1}(j), \boldsymbol{u_1}(b_1, l)) \in A_\varepsilon^n$; then generates $x_1$ i.i.d. according to $\prod_{i=1}^{n} p(x_{1i}|q_i, w_{1i}, u_{1i})$ and sends it.

The sender $TX_2$ to send $(b_2, b_3)$, knowing $\boldsymbol{q}$ and being non-causally aware of $\boldsymbol{w_1}(j), \boldsymbol{u_1}(b_1, l)$ finds a sequence $\boldsymbol{w_2}(m)$ in bin $b_2$ such that $(\boldsymbol{q}, \boldsymbol{w_1}(j), \boldsymbol{u_1}(b_1, l), \boldsymbol{w_2}(b_2, m)) \in A_\varepsilon^n$ and then finds a sequence $\boldsymbol{u_2}(k)$ in bin $b_3$ such that $(\boldsymbol{q}, \boldsymbol{w_1}(j), \boldsymbol{u_1}(b_1, l), \boldsymbol{w_2}(b_2, m), \boldsymbol{u_2}(b_3, k)) \in A_\varepsilon^n$. Finally, the sender $TX_2$ generates $x_2$ i.i.d. according to $\prod_{i=1}^{n} p(x_{2i}|q_i, w_{2i}, u_{2i})$ and sends it.

**Error probability analysis:** The messages are decoded based on strong joint typicality as in [8]. Assuming all messages to be equiprobable, we consider the situation where $(j = 1, b_1 = 1, b_2 = 1, b_3 = 1)$ was sent.

Here, we do the error analysis for the receiver $RX_2$. The decoding error proability for the receiver $RX_1$ can be evaluated in the same manner.

The receiver $RX_2$ upon receiving $\boldsymbol{y_2}$ and knowing $\boldsymbol{q}$ decodes $(j = 1, b_2 = 1, b_3 = 1)$ or $j(b_2, m)(b_3, k) = 1(1, m)(1, k)$ simultaneously [8]. We define the error event $E_{j(b_2,m)(b_3,k)}$ and $P_{e_2}^{(n)}$ as follows.

$E_{j(b_2,m)(b_3,k)} = \{(\boldsymbol{q}, \boldsymbol{w_1}(j), \boldsymbol{w_2}(b_2, m), \boldsymbol{u_2}(b_3, k), \boldsymbol{y_2}) \in A_\varepsilon^n\}$

$P_{e_2}^{(n)} = P\left(E_{1(1,m)(1,k)}^C \cup E_{j(b_2,m)(b_3,k) \neq 1(1,m)(1,k)}\right) \leq \varepsilon + \sum_{j\neq 1, b_2=b_3=1}^{\boxed{1}'} \cdots + \sum_{b_2\neq 1, j=b_3=1}^{\boxed{2}'} \cdots + \sum_{b_3\neq 1, j=b_2=1}^{\boxed{3}'} \cdots +$

$\sum_{j\neq 1, b_2\neq 1, b_3=1}^{\boxed{4}'} \cdots + \sum_{j\neq 1, b_3\neq 1, b_2=1}^{\boxed{5}'} \cdots + \sum_{b_3\neq 1, b_2\neq 1, j=1}^{\boxed{6}'} \cdots + \sum_{j\neq 1, b_3\neq 1, b_2\neq 1}^{\boxed{7}'} \cdots$

Let us choose $\boxed{1}', \boxed{3}', \boxed{6}'$ for evaluation. So, according to the codebook generation and the original distribution (8) in theorem 3, we have:

- $\sum_{j\neq 1, b_2=b_3=1}^{\boxed{1}'} \cdots \leq 2^{nT_1} \cdot p\left((\boldsymbol{q}, \boldsymbol{w_1}(j), \boldsymbol{w_2}(1, m), \boldsymbol{u_2}(1, k), \boldsymbol{y_2}) \in A_\varepsilon^n\right) \leq$

  $2^{nT_1} \|A_\varepsilon^n\| p(\boldsymbol{q})p(\boldsymbol{w_1}|\boldsymbol{q})p(\boldsymbol{w_2}|\boldsymbol{q})p(\boldsymbol{u_2}|\boldsymbol{q})p(\boldsymbol{y_2}|\boldsymbol{q}w_2u_2) \leq 2^{nT_1} \cdot 2^{nH(QU_2W_2W_1Y_2)} \cdot 2^{-nH(Q)} \cdot 2^{-nH(W_1|Q)} \cdot$

  $2^{-nH(W_2|Q)} \cdot 2^{-nH(U_2|Q)} \cdot 2^{-nH(Y_2|QU_2W_2)} = 2^{-n(I(U_2;W_2|Q)+I(W_2U_2;W_1|Q)+I(Y_2;W_1|QU_2W_2)-T_1)}$

- $\sum_{b_3 \neq 1, j=b_2=1}^{\boxed{3}'} \cdots \leq 2^{ns_2} \|A_\varepsilon^n\| p(q)p(w_1|q)p(w_2|q)p(u_2|q)p(y_2|qw_1w_2) \leq \cdots =$
$2^{-n(I(W_1;U_2W_2|Q)+I(Y_2;U_2|QW_1W_2)+I(U_2;W_2|Q)-s_2)}$

- $\sum_{j=1,b_3\neq 1,b_2\neq 1}^{\boxed{6}'} \cdots \leq 2^{n(s_2+t_2)} \|A_\varepsilon^n\| p(\cdots) \leq 2^{-n(I(U_2;W_2|Q)+I(W_1;W_2U_2|Q)+I(Y_2;U_2W_2|QW_1)-s_2-t_2)}$

Similarly, the other terms can be evaluated. In order to ($P_{e_2}^{(n)} \to 0$ as the block length $n \to \infty$), it is necessary and sufficient that:

$$\begin{cases} s_2 \leq I(W_1;U_2W_2|Q) + I(Y_2;U_2|QW_1W_2) + I(U_2;W_2|Q) \\ T_1 \leq I(U_2;W_2|Q) + I(W_2U_2;W_1|Q) + I(Y_2;W_1|QU_2W_2) \\ t_2 \leq I(U_2;W_2|Q) + I(W_1;W_2U_2|Q) + I(Y_2;W_2|QW_1U_2) \\ s_2 + T_1 \leq I(U_2;W_2|Q) + I(W_1;W_2U_2|Q) + I(Y_2;U_2W_1|QW_2) \\ s_2 + t_2 \leq I(U_2;W_2|Q) + I(W_1;W_2U_2|Q) + I(Y_2;U_2W_2|QW_1) \\ T_1 + t_2 \leq I(U_2;W_2|Q) + I(W_1;W_2U_2|Q) + I(Y_2;W_1W_2|QU_2) \\ T_1 + s_2 + t_2 \leq I(U_2;W_2|Q) + I(W_1;W_2U_2|Q) + I(Y_2;U_2W_1W_2|Q) \end{cases} \quad (\mathcal{A}_1),$$

from where, considering the binning conditions:
$I(W_2; W_1U_1|Q) \leq t_2 - T_2$ or $T_2 - t_2 \leq -I(W_2; W_1U_1|Q)$,
$I(U_2; U_1W_1W_2|Q) \leq s_2 - S_2$ or $S_2 - s_2 \leq -I(U_2; U_1W_1W_2|Q)$,
The relations ($\mathcal{A}_1$) yield to the constraints (10-8)-(10-14) in theorem 3. From a similar error analysis for the receiver $RX_1$, the inequalities (10-1)-(10-7) in theorem 3 are obtained.

## APPENDIX B

**Proof of theorem 4**

The proof is done in the same manner as in [16]. In (10-1)-(10-14) and $S_i \geq 0$, $T_i \geq 0$, $i = 1,2$, set $S_1 = R_1 - T_1$ and $S_2 = R_2 - T_2$. Then by eliminating $T_1$ and $T_2$, step by step, we reach to the following 37 inequalities (the bound constants $A_i$, $B_i$, $C_i$, $D_i$, $E_i$, $F_i$, $G_i$, $i = 1,2$ are the same as in theorem 3):

- $R_1 \leq D_1$,     $R_1 \leq C_2 + A_1$,     $R_1 \leq A_1 + B_1$,     $R_1 \leq G_1$,
$R_1 \leq F_2 + A_1$,     $R_1 \leq C_2 + E_1$,     $R_1 \leq B_1 + E_1$,     $R_1 \leq F_1 + A_1$,
$R_1 \leq F_1 + E_1$,     $R_1 \leq E_2 + A_1$,     $R_1 \leq F_2 + E_1$

- $R_2 \leq D_2$,     $R_2 \leq A_2 + C_1$,     $R_2 \leq A_2 + B_2$,     $R_2 \leq A_2 + E_1$

- $R_1 + R_2 \leq E_2 + E_1$,     $R_1 + R_2 \leq G_2 + A_1$,     $R_1 + R_2 \leq E_2 + A_1 + C_1$,
$R_1 + R_2 \leq E_2 + A_1 + B_2$,     $R_1 + R_2 \leq E_2 + A_1 + E_1$,     $R_1 + R_2 \leq G_1 + A_2$,
$R_1 + R_2 \leq C_2 + E_1 + A_2$,     $R_1 + R_2 \leq F_2 + A_1 + A_2$,     $R_1 + R_2 \leq B_1 + E_1 + A_2$,
$R_1 + R_2 \leq F_1 + A_1 + A_2$,     $R_1 + R_2 \leq G_2 + E_1$

- $2R_1 + R_2 \leq E_2 + A_1 + G_1$,     $2R_1 + R_2 \leq E_2 + A_1 + C_2 + E_1$,
$2R_1 + R_2 \leq E_2 + F_2 + 2A_1$,     $2R_1 + R_2 \leq E_2 + A_1 + B_1 + E_1$,
$2R_1 + R_2 \leq E_2 + F_1 + 2A_1$,

- $2R_2 + R_1 \leq F_1 + E_1 + 2A_2$,     $2R_2 + R_1 \leq A_2 + G_2 + E_1$,
$2R_2 + R_1 \leq F_2 + E_1 + 2A_2$,

- $3R_1 + 2R_2 \leq F_1 + E_1 + 2E_2 + 2A_1$,     $3R_1 + 2R_2 \leq F_2 + E_1 + 2E_2 + 2A_1$,

- $2R_1 + 2R_2 \leq G_2 + E_1 + E_2 + A_1$.

Then, by considering the information theoretic relations between the bound constants, we verify that 17 inequalities of the above 37 inequalities are redundant and finally we reach to twenty inequalities in theorem 4.

## APPENDIX C

Proof of theorem 5

It is sufficient to show that any element of $S_{CRC}^{Hod-1}(Z_1)$ for each $Z_1 \in \mathcal{P}_{CRC}^{Hod-1}$ is achievable. So, fix $Z = (QW_1W_2X_1X_2Y_1Y_2)$ and take any $(T_1, S_1, T_2, S_2)$ satisfying the constraints of the theorem.

**Codebook generation**: Consider $n > 0$, some distribution of the form (12) and

$$p(w_2|q) = \sum_{x_1,w_1} p(w_1|q)\, p(x_1|qw_1)\, p(w_2|x_1w_1q)$$

$$p(x_2w_2|q) = p(w_2|q)\, p(x_2|qw_2) = \sum_{x_1,w_1} p(w_1|q)\, p(x_1|qw_1)\, p(x_2w_2|x_1w_1q).$$

Therefore, by using random binning we can generate the sequences of $w_2$ and $x_2$ independently of $w_1$, $x_1$. So,

1. generate a n-sequence $q$, i.i.d. according to $\prod_{i=1}^{n} p(q_i)$,

2. for the codeword $q$, generate $\lfloor 2^{nT_1} \rfloor$ conditionally independent codewords $w_1(j), j \in \{1,2,\cdots,\lfloor 2^{nT_1} \rfloor\}$ according to $\prod_{i=1}^{n} p(w_{1i}|q_i)$,

3. for the codeword $q$ and each of the codewords $w_1(j)$, generate $\lfloor 2^{nS_1} \rfloor$ n-sequence $x_1(j,k), k \in \{1,2,\cdots,\lfloor 2^{nS_1} \rfloor\}$, i.i.d. according to $\prod_{i=1}^{n} p(x_{1i}|w_{1i}(j), q_i)$,

4. for the codeword $q$, generate $\lfloor 2^{nt_2} \rfloor$ n-sequence $w_2(l), l \in \{1,2,\cdots,\lfloor 2^{nt_2} \rfloor\}$, i.i.d. according to $\prod_{i=1}^{n} p(w_{2i}|q_i)$, and throw them randomly into $\lfloor 2^{nT_2} \rfloor$ bins such that the sequence $w_2(l)$ in bin $s_{21}$ is denoted as $w_2(s_{21}, l), s_{21} \in \{1,2,\cdots,\lfloor 2^{nT_2} \rfloor\}$,

5. for the codeword $q$ and each of the codewords $w_2(s_{21}, l)$, generate $\lfloor 2^{ns_2} \rfloor$ n-sequences $x_2(b,l)$, $b \in \{1,2,\cdots,\lfloor 2^{ns_2} \rfloor\}$, i.i.d. according to $\prod_{i=1}^{n} p(x_{2i}|w_{2i}(l), q_i)$, and throw them randomly into $\lfloor 2^{nS_2} \rfloor$ bins such that the sequence $x_2(b,l)$ in bin $s_{22}$ is denoted as $x_2(s_{22}, b, l), s_{22} \in \{1,2,\cdots,\lfloor 2^{nS_2} \rfloor\}$.

**Encoding**: The aim is to send a four dimensional message $(j, k, s_{21}, s_{22})$ whose first two components $j$ and $k$ are message indices and whose last two components $s_{21}$ and $s_{22}$ are bin indices. The messages actually sent over the genie-aided cognitive radio channel are $x_1$ and $x_2$. The message and bin indices are mapped into $x_1$ and $x_2$ as follows.

The sender TX1 to send $j$ and $k$ looks for $w_1(j), x_1(j,k)$ and sends $x_1(j,k)$.

The cognitive sender TX2 knowing $w_1(j), x_1(j,k)$ noncausally and $q$, to send $(s_{21}, s_{22})$ finds a sequence $w_2(l)$ in bin $s_{21}$ such that $(q, w_1(j), x_1(j,k), w_2(l)) \in A_\varepsilon^n$ and then finds a sequence $x_2(b,l)$ in bin $s_{22}$ such that $(q, w_1(j), x_1(j,k), w_2(l), x_2(b,l)) \in A_\varepsilon^n$ and sends $x_2(s_{22}, b, l)$.

**Decoding and analysis of error probability:** The receivers RX1 and RX2 decode the corresponding messages independently, based on strong joint typicality [8]. The inputs $x_1$ and $x_2$, to the genie-aided cognitive radio channel are received at the receivers as $y_1$ and $y_2$, according to the conditional distributions $p(y_1|x_1x_2)$ and $p(y_2|x_1x_2)$, respectively. It is assumed that all messages are equiprobable. Without loss of generality it is assumed that $(j = 1, k = 1; s_{21} = 1, s_{22} = 1)$ is sent with the codeword $q$, known to both receivers and senders. Notice that the first two components, $j$ and $k$, are message indices, whereas the last two components, $s_{21}$ and $s_{22}$, are bin indices. Now, we can find the constraints in theorem 1 such that the average probability of error $P_e^{(n)} \to 0$ as the block length $n \to \infty$.

The receiver $RX_1$, by receiving $y_1$ and knowing $q$, decodes $j = 1, k = 1; s_{21} = 1$ or $jk(s_{21}, l) = 11(1, l)$ simultaneously [8]. Therefore, we can define the event $E_{jk(s_{21},l)}$ and $P_e^{(n)}$ as follows, thereby establishing the constraints that lead to $(P_e^{(n)} \to 0$ as the block length $n \to \infty)$. (The state $j = k = 1$; $s_{21} \neq 1$ is not considered as error).

$$E_{jk(s_{21},l)} = \{(q, w_1(j), x_1(j,k), w_2(s_{21}, l), y_1) \in A_\varepsilon^n\}$$

$$P_e^{(n)} = P\left\{E_{11(1,l)}^c \cup E_{jk(s_{21},l) \neq 11(1,\hat{l})}\right\} \leq P(E_{11(1,l)}^c) + \sum_{jk(s_{21},l) \neq 11(1,l)} P(E_{jk(s_{21},l)}) \leq \varepsilon + \overset{\boxed{1}}{\underset{j \neq 1, k = s_{21} = 1}{\sum}} \cdots +$$

$$\overset{\boxed{2}}{\underset{j = s_{21} = 1, k \neq 1}{\sum}} \cdots + \overset{\boxed{3}}{\underset{j \neq 1, k \neq 1, s_{21} = 1}{\sum}} \cdots + \overset{\boxed{4}}{\underset{j \neq 1, s_{21} \neq 1, k = 1}{\sum}} \cdots + \overset{\boxed{5}}{\underset{k \neq 1, s_{21} \neq 1, j = 1}{\sum}} \cdots + \overset{\boxed{6}}{\underset{j \neq 1, k \neq 1, s_{21} \neq 1}{\sum}} \cdots \ ,$$

from where in order to $(P_e^{(n)} \to 0$ as the block length $n \to \infty)$, in accordance with the codebook we reach to four necessary and sufficient conditions for $S_1$, $T_1$, $t_2$, $s_2$ (the other two resulted inequalities are redundant) and then, by considering the binning conditions, the inequalities (13-1)-(13-4) are obtained.

Similarly, for the receiver $RX2$, we can define the similar event and $P_e^{(n)}$, thereby establishing (13-5)-(13-8).

# APPENDIX D

**Proof of theorem 6**

The proof parallels that of theorem 4. By setting $S_1 = R_1 - T_1$ and $S_2 = R_2 - T_2$, in (13-1)-(13-8) and $S_i \geq 0$, $T_i \geq 0$, $i = 1,2$, eliminating $T_1$ and $T_2$; removing redundant inequalities (for brevity, the details are omitted), we find (15-1)-(15-11)